\documentstyle[12pt]{article}

         \def\be{\begin{equation}}
         \def\bea{\begin{eqnarray}}

         \def\ee{\epsilon}

         \def\ee{\end{equation}}
         \def\eea{\end{eqnarray}}
         \def\R{\rm {I\kern-.200em R}}
         \def\C{\rm {I\kern-.520em C}}
\begin{titlepage}
\begin{document}
\vspace*{5mm}
\begin{center} {\Large \bf Laughlin States on the Poincare half-plane\\
\vskip 0.35cm
 and its quantum group symmetry}\\
\vskip 1cm
M.Alimohammadi and H.Mohseni Sadjadi\\
\vskip 1cm
{\it Physics Department, University of Teheran, North Karegar,} \\
{\it Tehran, Iran }\\
{\it Institute for studies in Theoretical Physics and Mathematics, }\\
{\it P.O.Box 19395-5746, Tehran, Iran}\\
\end{center}
\vskip 2cm
\begin{abstract}

We find the Laughlin states of the electrons on the Poincare half-plane in
different representations. In each case we show that there exist a
quantum group $su_q(2)$ symmetry such that the Laughlin states are a representation of
it. We calculate the corresponding filling factor by using the plasma analogy of
the FQHE.
\end{abstract}
\newpage
{\bf 1- Introduction}

\hskip 0.25cm

Studying the behaviour of the charged particles on the two-dimensional surface
in the presence of the strong magnetic field has led to the discovery of the
fractional quantum Hall effect (FQHE) [1,2]. To explain this phenomenon, Laughlin
proposed a suitable $N$-particle wave function which describes the FQHE of the
filling factor $\nu={1\over m}$, where $m$ is an odd integer [3]. Laughlin's
model
has also a beautiful analogy with an incompressible fluid of interacting plasma.

After that, the quantum mechanics of the nonrelativistic particles in a
uniform magnetic field was studied for different two-dimensional surfaces. The
first was the sphere on which the magnetic field was produced by a magnetic monopole[4].
Recently the topological torus[5] and arbitrary two-dimensional compact
Riemann surfaces were studied [6].

One of the important point in the physics of the fractional quantum Hall effect
is to understand the incompressibility feature of this problem in the language
of the symmetries of this theory. In Ref.[7], it is shown that this feature
relates to the exsistance of the Fairlie-Fletcher-Zachos (FFZ) algebra [8] as
a symmetry algebra of the Hamiltonian. As this algebra reduces to the
area-preserving diffeomorphism, it can explain the incompressibility.
It was also shown that the generators of the FFZ algebra, which are the magnetic
translation operators, could represent the $su_q(2)$ algebra where $q$ is a
function of the magnetic field[5,7,9].

The case of non-compact surfaces,
and in special the upper half-plane with Poincare metric was also studied in
several papers [10,11,12]. In these articles the one-particle wave functions
and the symmetries of the Hamiltonian were discussed. In Ref.[13] we began our
investigation of $su_q(2)$ symmetry for this surface by finding the
generators of this quantum algebra and showing that the one-particle ground
state is a representation of this $su_q(2)$.

In this paper we are going to complete our study about the FQHE on the Poincare
half-plane by calculating the Laughlin states. We will find different
representations of this state. To clarify what we mean by different representations,
we remind the reader that in the original work of Laughlin, the ground states were the
eigenstates of the angular momentum. But in our case the angular momentum is
not the symmetry of the Hamiltonian, nevertheles there are three operators which
commute with the Hamiltonian and generate the $SL(2,R)$ algebra. By
different Laughlin states we mean that we will find the Laughlin wave functions
which are simultaneous eigenfunctions of Hamiltonian and different symmetry
operators. In all cases we will
show that the Laughlin
states form a representation of $su_q(2)$ .

We will also discuss the filling
factor. The calculation of the filling factor (which is defined as the ratio of
the total number of the electrons to the degeneracy of the first Landau level)
is not clear in the non-compact surface. This is because the degeneracy and also
the total area are both infinite in this case. Therefore we must calculate it
in a different way. As it will be seen, we will compute $\nu$ by using the
plasma analogy.

In section 2 we will write the Laughlin states in such a way that it will be
the eigenstates of the operators ${\cal L}_1^{-1} {\cal L}_2$ which was used in
Ref.[12]. In
section 3 another symmetry operator will be used (the operator ${\cal L}_2$ which
generates dilation) and the single-particle and also the Laughlin wave functions
will be found. By caculating the effective interaction potential, we will find
the corresponding filling factor of these states. The generators of the quantum
group symmetry with $B$-dependent $q$ will be also found. The degeneracy of the
the first Landau level which will be considered in sections one and two are
infinite and the states are labelled by a continuous parameter. For completeness of our study,
we will consider the discrete degenerate states in section 3.

\hskip 1cm

{\bf 2- Laughlin states as eigenstate of ${\cal L}_1^{-1} {\cal L}_2$}

\hskip 0.25cm

Consider the upper half-plane $\{ z=x+iy, y>0 \}$ with the metric:

\be ds^2 = {{d x^2 + d y^2}\over {y^2}} \ee
For a covariently constant magnetic field $B$ and in the symmetric gauge;
$A_z=A_{\bar z}={B\over {2y}}$, the one particle Hamiltonian is [11,13]:

\be H=-y^2 \partial {\bar \partial }  +{{iB}\over2} y (\partial +{\bar \partial})
+B^2/4\ee
(We take the electron mass $m=2$). The symmetry operators of this
Hamiltonian are:

$$L_1=\partial_x=\partial +{\bar \partial }$$
\be L_2=x\partial_x+y\partial _y =z\partial +{\bar z}{\bar \partial }\ee
$$L_3=(y^2 -x^2)\partial _x-2xy \partial _y -2iBy $$
The operators $L_i$ generates the $SL(2,R)$ algebra. The ground states with
energy $B/4$ are [13]:

\be \psi_0(z,{\bar z})=y^B f(z) , \ee
where $f(z)$ is an arbitrary holomorphic function.
In Ref.[13] it was shown that if we demand that $\psi _0 (z,{\bar z})$ be an
eigenfunction of $b=L_1^{-1}L_2$ with eigenvalue $\lambda$ \footnote {By solving
the eigenvalue problem  $L_1^{-1}L_2\psi = \lambda \psi$ , we mean solving the
equation $(L_2-\lambda L_1  )\psi =0$ .}
, it takes the form:

\be \psi_0 (\lambda |z,{\bar z})=y^B (\lambda -z)^{-B}\ee

If we define $T_\xi =exp (\xi_1 c+\xi _2 b)$ , where $c=L_1$ , then it was
shown that the operators:

\be J_+ = {{T_{\vec \xi} -T_{\vec \eta}}\over {q-q^{-1}}} \hskip 2mm,
\hskip 2mm J_- = {{T_{- \vec \xi} -T_{-\vec \eta}}\over {q-q^{-1}}}\hskip 2mm,
\hskip 2mm q^{2J_0}=T_{\vec \xi -\vec \eta}\ee
satisfy the $su_q(2)$ algebra [14]

\be [J_0 , J_{\pm}] =\pm J_{\pm} \ee
$$[J_+,J_-]={1\over {q-q^{-1}}} (q ^{2J_0} -q^{-2J_0}), $$
and $\psi _0 (\lambda |z, \bar z)$ are a representation of this algebra. In eq.(6)
$\vec \xi =(\xi_1,\xi_2)$ and $\vec \eta =(\xi_1 , -\xi_2)$.

Now to construct the $N$-particle wave function we assume the magnetic field to
be so strong so that we can approximately neglect the electron-electron intractions.
In this case the Laughlin wave function takes the form:

\be \psi _m (z_i, \bar z_i )=\prod ^ N_{j<k} (z_j -z_k)^ m f(z_1,...,\bar z_N) .\ee
We will take $f(z_i,\bar z_i)$ to be totaly symmetric under the interchange $z_i \leftrightarrow z_j$
so that , with $m$ an odd positive integer , $\psi _m$ will be totaly antisymmetric.
$f(z_i,\bar z_i)$ must be found such that $\psi_m$ will be the ground state wave
function of the noninteracting Hamiltonian $H= \sum ^N_{i=1} H_i$,
where $H_i$ is
defined as in eq.(2), with energy $NB/4$. In this way it can be seen
that $f(z_i,\bar z_i)$ is:

\be f(z_1,...,\bar z_N)=\prod ^N_{i=1} y_i^B \psi(z_1,..., z_N)
e^{\lambda_1 \bar z_1  + ... +\lambda _N \bar z_n} \ee
with $\sum ^N _{i=1} \lambda _i=0$. The condition of symmetrization of $f(z_i,\bar z_i)$,
forces us to take
all $\lambda _i$ equal and therefore $\lambda _i=0$ , and $\psi (z_1,...,z_N)$ equal
$\prod ^N _{i=1}\psi _0 (z_i)$. So:

\be \psi _m (z_i, \bar z_i )=\prod ^N_{j<k} (z_j -z_k)^ m \prod ^N_{i=1}
y_i ^B \psi _0 (z_i).\ee
Now we will determine $\psi _0 (z_i)$ such that $\psi _m$ will be an eigenfunction
of ${\cal L}_1^{-1}{\cal L}_2$ with eigenvalue $\lambda$. ${\cal L}_1$ and ${\cal L}_2$ are:

\be {\cal L}_1=\sum ^N_{i=1} (\partial _i +\bar \partial _i)\ee
$${\cal L}_2=\sum ^N_{i=1} (z_i \partial _i +\bar z_i \bar \partial _i).$$
By using the following relations:

\be {\cal L}_1 \prod ^N _{j<k} (z_j -z_k)^m =0 \ee
$$ {\cal L}_2 \prod ^N _{j<k} (z_j -z_k)^m = {{mN(N-1)}\over 2}\prod^N_{j<k}
(z_j -z_k)^m$$
and by using the condition of symmetrization of $\psi (z_1,...,z_N)$ , one obtains:
\be \psi _m (\lambda, z_i, \bar z_i )=\prod ^N_{j<k} (z_j -z_k)^ m \prod ^N_{i=1}
y_i^B(\lambda -z_i)^{-B-{{m(N-1)}\over 2}}\ee
It can be seen that for $N=1,\psi _m$ reduces to eq.(5). Also it can be checked
that the above states fom an infinite dimensional representation of $su_q(2)$ algebra:
\be J_\pm \psi _m (\lambda, z_i, \bar z_i )=[1/2 \mp {\lambda/\xi_1}]_q \psi_m
(\lambda \mp \xi_1,z_i,\bar z_i)\ee
$$ q^{\pm J_0}\psi _m (\lambda, z_i, \bar z_i )=q^{\mp \lambda/\xi_1}\psi_m
(\lambda, z_i, \bar z_i)$$
where $[x]_q$ is defined by $[x]_q=(q^x-q^{-x})/(q-q^{-1})$ and $J_\pm$ and ${J_0}$
are defined in the same way as eq.(6), with $T_{\bar \xi}=exp(\xi_1 {\cal L}_1
+\xi_2 {\cal L}_1^{-1}{\cal L}_2)$.

For better understanding of the physics behind the Laughlin states, we will
find another representation of the Laughlin states which is more suitable.

\hskip 1cm

{\bf 3- Laughlin wavefunction as eigenstates of ${\cal L}_2$}

\hskip 0.25cm

Let us first consider the one-particle wavefunction. If we demand that the state
(4) be an eigenstate of the operator $L_2$ with eigenvalue $m$ , it can
easily be found:
\be |m>= \psi _m (z, \bar z)=y^B z^{m-B} \ee
These states form a representation of the quantum group $su_q(2)$ .This can be
seen as follows: define $E^\pm $ and $k$ as:
\be E^+=-z [L_2+\alpha+ \beta]_q \hskip 2mm, \hskip 2mm E^-=z^{-1} [L_2+\alpha-
\beta]_q \hskip 2mm, \hskip 2mm k=q^{L_2+\alpha} \ee
Then we can verify that:
\be [E^+ , E^-] |m>={{k^2-k^{-2}}\over {q-q^{-1}}}|m>\hskip 4mm, \hskip 4mm kE^\pm
k^{-1} |m>=q^\pm E^\pm |m>\ee

For the $N$-particle state we can see that under the same assumptions of
the last section, the following wave function:

\be \psi _m (z_i, \bar z_i )=\prod ^N_{i=1}y_i^B \prod ^N _{i<j} (z_i -z_j)
^ {m - B}
\ee
is: a) eigenstate of $H=\sum H_i$ with eigenvalue $NB/4$ , b) eigenstate of
${\cal L}={2\over{N(N-1)}}({\cal L}_2+{{NB(N-3)}\over 2})$
with eigenvalue $m$ and c) totaly antisymmetric.

The generators of $su_q(2)$ are now:
\be E^+= -\prod ^N_{i<j} (z_i-z_j)[{\cal L}+\alpha+ \beta]_q \ee
$$ E^-= \prod ^N_{i<j} (z_i-z_j)^{-1}[{\cal L}+\alpha- \beta]_q $$
$$ k=q^{{\cal L}+\alpha} $$
To ensure the Fermi-Dirac statistic for the wavefunction (18), $m-B$ must be
1,3,5,... . Since $E^-$ is lowering operator , and reduces $m$ by $1$ , there
should exist a lowest state $|m_{min}>$ :
                        \be E^-|m_{min}>=0 \ee
This condition can determine the deformation parameter $q$ as (by choosing $a=\beta$)
                        \be q=exp ({{\pi i}\over {B+1}}) \ee
This equation relates $q$ to the magnetic field as in the cases of the plane [7] ,
sphere [9] and torus [5] .

To calculate the filling factor that corresponds to the Laughlin state (18),
we proceed to the same method that was followed by Laughlin [3] , that is we
introduce the effective classical potential energy $\phi$ in $|\psi_m|^2=e^{-\beta \phi}$. If we
set the arbitrary effective temperature ${1\over \beta}$ equal to $m-B$, we find:

\be \phi = -(m-B)^2 \sum ^N_{i<j} \ln |z_i-z_j|^2  -(m-B)\sum _i \ln y_i ^{2B}\ee
The first term is the natural coulomb interaction of the particles with charge
$m-B$ . This is because the solution of the Laplace equation in the Poincare
half-plane is logarithmic. If one calculates the Laplace-Beltrami operator
for the metric (1) , one finds :
\be \nabla^2 \phi = {1\over {\sqrt g}}\partial_i \sqrt g
g ^{ij} \partial _j \phi=y^2 (\partial ^2 _x+\partial^2_y)\phi \ee
Then
 \be \nabla^2 \ln z \bar z=\Delta ({\bf r})= y^2 \delta (x) \delta (y) \ee
where $\Delta({\bf r})$  is the delta function on the Poincare half-plane :
\be \int \Delta ({\bf r}) \sqrt g dx dy =1 \ee
The second term of eq.(22) is the interaction of these particles with the uniform
neutralizing background of charge density $\rho_0={{-B}\over {2\pi}}$ ;
\be \nabla ^2(-\ln y^{2B})=-4 \pi ({{-B}\over {2\pi}}). \ee
But the plasma must be electrically neutral everywhere, so the total charge of
these particles must be equal to the background charge , and
this leads to the charge density:
                        \be \rho_m={N\over A}={{\rho_0}\over {m-B}}, \ee
where $N$ is the total number of the charged particles and $A$ is total area.
Therefore the filling factor $\nu = {{\rho_m}\over {\rho_0}}$ is equal to:
                        \be \nu ={1\over {m-B}}.\ee
So the wavefunction (18) corresponds to filling factor $\nu={1\over M}$ where
$M=m-B$ is a
positive odd integer.

\hskip 1cm

{\bf 4- Discrete representation}

\hskip 0.25cm

To make complete our study of the FQHE on Poincare half-plane, we are going to
discuss the Laughlin state as a discrete representation of $su(1,1)$ algebra.
As discussed in Ref.[11] , if we define :
   $$ J_0=-{i\over 2}(L_1-L_3)$$
\be J_1=  -{i\over 2}(L_1+L_3)\ee
$$J_2=-i L_2$$
then it can be seen that they satisfy the $su(1,1)$ algebra:
$$ [J_0, J_1]=iJ_2 $$
\be  [J_0, J_2]=-iJ_1\ee
$$ [J_1, J_2]=-iJ_0 $$
and the Hamiltonian (2) becomes the Casimir , $C=J_0^2 -J_1^2 -J_2^2=-4H+B^2$.
This algebra have two kinds of representation, the discrete and continuous.
These representations are labelled by the eigenvalues of the Casimir operator
and the compact operator $J_0$
\pagebreak
$$C|j,n>=j(j+1)|j,n>$$
\be J_0|j,n>=n|j,n> \ee
$$<jn'|jn>=\delta _{nn'}$$

The unitary irreducible representation of the discrete series is divided to
two kind $D_j^+$ or $D_j^-$, depending on the values of $j$ . for $j>0$ :
                        \be D_j^+ = \{|j,j+1>, |j,j+2>, ...\} \ee
with $J_- |j,j+1>=0$ , and for $j<0$ :
                        \be D_j^- = \{|j,j>, |j,j-1>, ...\} \ee
with $J_+ |j,j>=0$ . $J_\pm$ are as usual ; $J_1\pm iJ_2$.

Now if we choose the eigenstates of the Hamiltonian to be in the discrete
series, then $j$ takes the values $-B+n$ where $n=0,1,2,...$ . The ground states
correspond to $j=-B$ and therefore we are in the $D_j^-$ series. We have infinite
discrete degenerate ground states :
\be |-B,-B> \ \ , \ \  |-B, -B-1>, ... \ee
To find these states explicitly, we choose the ground state (4) to be the
eigenstates of $J_0$ with eigenvalue $n$ . By some calculation, we find:
\be \psi_n (z, \bar z)=y^B {{(z-i)^{n-B}}\over {(z+i)^{n+B}}}\ee

The quantum group generators are the same as those in eq.(16), by replacing $z$
in eq.(16) with ${{z-i}\over {z+i}}$ and $L_2$ with $J_0$.
Also as our states are those in
eq.(34), so $n_{max}=-B$ and therefore $E^+|n_{max}>=0$ which gives $q$
(by choosing $\alpha=\beta=-1/2$)
as:
\be q=exp ({{\pi i}\over {B+1}}) \ee

By the same reasoning, the $N$-particle wave functions is $\psi _m (z_i,\bar z_i)$ in eq.(10) ,
where we determine $\psi _o(z_i)$ such that the $\psi _m (z_i,\bar z_i)$ will be the eigenfunction of
$J=\sum_i J_0^i$ with eigenvalue $M$. A lengthy calculation shows that:
\be \psi_m ^M (z_i, \bar z_i)=\prod ^N_{j<k} (z_j-z_k)^m \prod ^N_{j=1}y^B_j
{{(z_j-i)^{M/N-B-m(N-1)/2}}\over{(z_j+i)^{M/N+B+m(N-1)/2}}}\ee
with $m=2k+1$ . Finaly the suitable $su_q(2)$ generators are:
\pagebreak
\be E^+= -\prod ^N_{j=1} ({{z_j-i}\over {z_j+i}})^{1/N}[J+\alpha+ \beta]_q \ee
$$ E^-= \prod ^N_{j=1} ({{z_j-i}\over {z_j+i}})^{-1/N}[J+\alpha- \beta]_q $$

$$ k=q^{J+\alpha} $$

\hskip 1cm

{\bf 5- Conclusion}

\hskip 0.25cm

As mentioned in the introduction, one way to describe the behaviour of the
electron in the FQHE is the concept of incompressible fluid, and its presence
can be seen by checking the existance of the quantum group symmetry of the
Laughlin states. In this paper we showed that in all cases, there are such
symmetries and therefore we believe that this indicates that the collective
motion of the electrons in FQHE on the Poincare half-plane are also
incompressible.

The last point is that, it can be easily shown that the operator :
\be L=g_1(z)L_1+g_2(z)L_2+g_3(z)L_3+g_4(z)\ee
with arbitrary holomorphic functions $g_i(z)$'s, commutes with the Hamiltonian at
the level of the ground state :
\be [H,L]\psi _0 (z,\bar z)=0 \ee
It can be shown that one can write the $N$-particle wavefunction to be the
eigenstate of $L$, and with suitable choosing of $g_i (z)$ , these functions can
be made normalizable. The importance of this point will appear when we consider that
the wavefunctions of the previous sections are not normalizable.

\hskip 1cm

{\bf Acknowledgement}

\hskip 0.25cm

We would like to thanks S.Rouhani and V.Karimipour for their careful reading
of this article.
We would also like to thanks the research vice-chancellor of the university of Tehran.

\end{titlepage}
\end{document}